\documentclass[prc,twocolumn,floatfix,groupedaddress,nofootinbib,preprintnumbers,amsmath,amssymb,amsfonts,superscriptaddress,widetable,href]{revtex4}
\usepackage{amsmath,amssymb,amsfonts}
\usepackage{hyperref}
\usepackage{ulem}
\usepackage{graphicx}
\usepackage{color}
\usepackage{longtable}

\newcommand{\be}{\begin{equation}}
\newcommand{\ee}{\end{equation}}
\newcommand{\bea}{\begin{eqnarray}}
\newcommand{\eea}{\end{eqnarray}}

\allowdisplaybreaks

\begin{document}

\title{Interdependence of different symmetry energy elements} 
\author{C. Mondal}
\email {chiranjib.mondal@saha.ac.in}
\address{Saha Institute of Nuclear Physics, 1/AF Bidhannagar, Kolkata
{\sl 700064}, India}
\address{Homi Bhabha National Institute, Anushakti Nagar, Mumbai 400094, India.}
\author{B. K. Agrawal}
\email {bijay.agrawal@saha.ac.in}
\address{Saha Institute of Nuclear Physics, 1/AF Bidhannagar, Kolkata
{\sl 700064}, India}
\address{Homi Bhabha National Institute, Anushakti Nagar, Mumbai 400094, India.}
\author{J. N. De}
\email {jn.de@saha.ac.in}
\address{Saha Institute of Nuclear Physics, 1/AF Bidhannagar, Kolkata
{\sl 700064}, India}
\author{S. K. Samaddar}
\email {santosh.samaddar@saha.ac.in}
\address{Saha Institute of Nuclear Physics, 1/AF Bidhannagar, Kolkata
{\sl 700064}, India}
\author{M. Centelles}
\email {mariocentelles@ub.edu}
\address{Departament de F\'isica Qu\`antica i Astrof\'isica and Institut de 
Ci\`encies del Cosmos (ICCUB), Facultat de F\'isica, Universitat de Barcelona, Mart\'i 
i Franqu\`es 1, E-08028 Barcelona, Spain}
\author{X. Vi\~nas}
\email {xvinasg@ub.edu}
\address{Departament de F\'isica Qu\`antica i Astrof\'isica and Institut de 
Ci\`encies del Cosmos (ICCUB), Facultat de F\'isica, Universitat de Barcelona, Mart\'i 
i Franqu\`es 1, E-08028 Barcelona, Spain}

\date{\today}

\begin{abstract}
 Relations between the nuclear symmetry energy coefficient and its
density derivatives are derived. The relations hold for a class 
of interactions with quadratic momentum dependence and a power-law density 
dependence. The structural connection between the different
symmetry energy elements as obtained seems to be followed by almost all
reasonable nuclear energy density functionals, both relativistic and 
non-relativistic, suggesting a universality
in the correlation structure. This, coupled with known values of some
well-accepted  constants related to nuclear matter, helps in  
constraining values of different density derivatives of the nuclear symmetry energy
shedding light on the isovector part of the nuclear interaction.
\end{abstract}

\maketitle

The nuclear symmetry energy elements encode in them imprints of the
nature of the isovector part of the nuclear interaction.  They have
profound implications not only in terrestrial nuclear physics but
also in astrophysics and cosmology \cite{Woods97,
Lattimer07,Li15a}.  Some of them like the symmetry energy coefficient
$C_2(\rho)$ \Big($=\frac{1}{2}\left(\frac{\partial^2 e(\rho,\delta)}
{\partial \delta^2}\right)_{\delta =0}$, where $e(\rho, \delta)$ is
the energy per nucleon of nuclear matter of density $\rho$ at isospin asymmetry
$\delta$\Big) is now  known in tighter bounds at the saturation  density
$\rho_0$ \cite{Moller12,Jiang12} of symmetric nuclear matter (SNM). 
From analysis of the giant dipole resonance (GDR) of $^{208}$Pb, a 
well-constrained estimate of $C_2(\rho)$ at a somewhat lesser density 
($\rho =0.1$ fm$^{-3}$) \cite{Trippa08} is also known. The
value of the density slope of the symmetry energy $L_0$ ($=L(\rho_0)=3\rho_0
\left(\frac{\partial C_2}{\partial \rho}\right)_{\rho_0}$) is less
certain \cite{Centelles09,Agrawal12, Li15a}.  The shroud of uncertainty
looms even larger on the higher derivatives of the symmetry energy [e.g.,
$K_{sym}^0$ ($=K_{sym}(\rho_0)=9\rho_0^2 \left(\frac{\partial^2C_2}{\partial
\rho^2}\right)_{\rho_0}$) or $Q_{sym}^0$ ($=Q_{sym}(\rho_0)=27\rho_0^3
\left(\frac{\partial^3C_2}{\partial \rho^3}\right)_{\rho_0}$)] and
on the difference between the neutron and proton  effective masses
$\Delta m^*_0$ [=$(m_n^*-m_p^*)/m$] in neutron-rich matter at $\rho_0$.
The values of $K_{sym}^0$ and $Q_{sym}^0$, in different parametrizations
of the Skyrme energy density functional (EDF) lie in very wide ranges
[$-700$ MeV $< K_{sym}^0 < $ 400 MeV; $-800$ MeV $ < Q_{sym}^0 < 1500 $ MeV ]
\cite{Dutra12, Dutra14} whereas there are divergent predictions on the
value of $\Delta m^*_0$ from theoretical studies based on microscopic
many-body theories \cite{Zuo05,Dalen05} or phenomenological approaches
\cite{Ou11,Sellahewa14,Chen09,Kong17}.  Such large uncertainties belie
a satisfactory understanding of the nuclear isovector interaction.

There is a sliver of expectation that the entities
$C_2^0$ $(=C_2(\rho_0))$, $L_0$, $K_{sym}^0$, etc. may have an intrinsic 
correlation among them. 
Finding a correlated structure for these symmetry energy elements
helps in making a somewhat more precision statement on an otherwise
uncertain isovector indicator as it may be tied up to another one  
known with more certainty. There is a hint of a relatively weak positive 
correlation between
$C_2^0$ and $L_0$ \cite{Dutra12, Dutra14}.  From observation of the
computed values of $L_0$ and $K_{sym}^0$ with selected sets of 
non-relativistic and
relativistic EDFs, 
a fairly linear relationship between $K_{sym}^0$
and $L_0$ is also suggested 
\cite{Yoshida06,Danielewicz09,Chen09,Vidana09,Ducoin11,Dong12,Santos14}. 
The present communication aims to point to some possible 
universal aspects of the nuclear EDFs related to nuclear symmetry energy elements.  
As we will see, in a general mean-field approach analytic relationships can be built 
up among the nuclear symmetry energy elements tieing them in a correlated 
structure and helping in finding their values in terms of a few empirical 
nuclear constants.

For symmetric nuclear matter at density $\rho$, with energy density ${\cal 
H}$, and at zero temperature ($T=0$), the chemical potential of the nucleon is given by 
\bea
\label{chem1}
\mu\;=\;\mathcal E_F\;=\;\frac{P_F^2}{2m^*}+V\;=\;\frac{P_F^2}{2m}+U,
\eea
where $\mathcal E_F$ is the Fermi energy, $P_F$ is the Fermi momentum, and the effective 
mass $m^*$ and the single-particle potential $V$ are given, respectively, by
$\hbar^2/2m^* = \delta {\cal H} / \delta {\mathcal K}$ and $V= \delta {\cal H} / \delta \rho$, 
where $\frac{\hbar^2}{2m}\mathcal K$ is the kinetic energy density. One also can redefine the   
single-particle potential as $U$ by including within it the effective mass contribution, 
as done in the r.h.s. of Eq.~(\ref{chem1}). We make no special assumption about the 
nucleonic interaction except that it is density dependent to simulate many-body
forces and that it depends quadratically on the momentum;  
thus, the single-particle potential $U$ separates into three parts, 
\bea
\label{sp1}
U=V_0+P_F^2 V_1+V_2.
\eea
The term $(V_0+P_F^2 V_1)$ on the right is the Hartree-Fock potential and
the last term $V_2$ is the rearrangement potential that arises from 
the density dependence in the interaction. The term $V_1$ 
comes from the momentum dependence:
\bea
\label{mstar5}
\frac{P_F^2}{2m^*}=\frac{P_F^2}{2m}+P_F^2 V_1.
\eea
In general, $m^*$ is momentum and energy dependent, in the 
mean-field level the energy dependence is ignored and the 
momentum dependence is taken at the Fermi surface. 
The rearrangement energy does not enter explicitly in the energy 
expression when written in terms of the mean-field potential 
\cite{Brueckner59,Bandyopadhyay90}, the energy per nucleon for SNM at 
density $\rho$ is then given by, 
\bea
\label{eners}
e&=&\frac{1}{2m}\left<p^2\right>+\frac{1}{2}\left<p^2\right>V_1+\frac{1}{2}V_0\nonumber\\
&=&\frac{1}{4}\left(\frac{1}{m}+\frac{1}{m^*}\right)\left<p^2\right>+\frac{1}{2}V_0.
\eea
The Gibbs-Duhem relation relates the chemical potential and energy as,
\bea
\label{gibbs}
\mu =e+\frac{P}{\rho},
\eea
where $P$ is the pressure of the system. At zero pressure this leads to 
the Hugenholtz-Van Hove theorem \cite{Hugenholtz58} which has recently been 
used to link nucleon single-particle characteristics to macroscopic isovector 
properties in Ref. \cite{Chen12}. Keeping this in mind, from 
Eqs. (\ref{chem1}), (\ref{eners}) and (\ref{gibbs}), 
the energy per nucleon for SNM can be written as \cite{De15},
\bea
\label{ener1}
e=\frac{P_F^2}{10m}\left(3-2\frac{m}{m^*}\right) -V_2+\frac{P}{\rho },
\eea
where $\langle p^2 \rangle= 3 P_F^2/5$ has been used.

For asymmetric nuclear matter,  the equation for the energy per nucleon
can be generalized  as
\bea
\label{ener4}
e(\rho,\delta)&=&\frac{1}{\rho}\left[\sum_{\tau}\frac{P_{F,\tau}^2}{10m}\rho_{\tau}\left(
3-2\frac{m}{m_{\tau}^*(\rho)}\right)\right]-V_2(\rho,\delta)\nonumber\\
&&+\frac{P(\rho,\delta)}{\rho}.
\eea
In Eq.(\ref{ener4}), $\tau $ is the isospin index, 
$\rho_{\tau}=(1+\tau \delta )\rho /2$; 
here, $\tau =1$ for neutrons and $\tau =-1$ for protons. 
The Fermi momentum for the individual species can be written as
$P_{F,\tau}=g_2\rho_{\tau}^{1/3}$ 
with $g_2=( 3\pi^2)^{1/3}\hbar$.
The density-dependent nucleon effective mass 
is written as 
\bea
\label{mstar6}
\frac{m}{m_{\tau}^*(\rho)}=1+\frac{k_+}{2}\rho+\frac{k_-}{2}\rho \tau \delta.
\eea
This is a generalization from $\frac{m}{m^*}=1+k\rho $, usually taken in a nonrelativistic
prescription for SNM \cite{Bohr69}.
The density dependence in the rearrangement potential is taken as
\bea
\label{vrea}
V_2(\rho,\delta)=(a+b\delta^2)\rho^{\tilde\alpha} ,
\eea
which is independent of the isospin index $\tau$.
The constant $a$ weighs the rearrangement potential for SNM, whereas the constant $b$ 
is a measure of the asymmetry dependence of the rearrangement potential. 

The energy per nucleon $e(\rho,\delta)$ can also be
written  in terms of the symmetry 
energy coefficients as 
\bea
\label{ener5}
e(\rho,\delta)= e(\rho,\delta=0)+C_2(\rho )\delta^2+C_4(\rho )\delta^4+\cdots
\eea
The expression for the  pressure $P(\rho,\delta )=
\rho^2\frac{\partial e}{\partial \rho}$ follows  from the
above equation. Hence, the right hand side of Eq.(\ref{ener4}) can be expanded in 
powers of
$\delta $ using the expressions for $P(\rho,\delta )$
and $V_2(\rho,\delta )$ and using Eq.(\ref{mstar6}). Comparing then  with Eq.(\ref{ener5}) and equating
coefficients
of the same order in $\delta$, one gets the expression for $C_2(\rho)$ as 
\bea
\label{c2r1}
C_2(\rho)&=&-b\rho^{\tilde\alpha}+\rho\frac{\partial C_2(\rho)}{\partial \rho}+y\rho^{2/3}\left[
-\frac{5}{3}k_-\rho\right.\nonumber\\&&\left.+\frac{5}{9}(1-k_+\rho)\right],
\eea
with $y=\frac{g_2^2}{10m}\frac{1}{2^{2/3}}$. The relation between 
$C_2(\rho)$ and its density derivative is a direct consequence of the 
Gibbs-Duhem relation.
At saturation 
density  the symmetry energy coefficient $C_2$  reads as, 
\bea
\label{c2r2}
C_2^0&=&-b\rho_0^{\tilde\alpha}+\frac{L_0}{3}+E_F^0\left[-\frac{1}{3}k_-\rho_0
\right.\nonumber\\
&&\left.+\frac{1}{9}(1-k_+\rho_0)\right],
\eea
where $E_F^0=5y\rho_0^{2/3}$ is the Fermi 
energy at $\rho_0$. 
Similar equations can be obtained for higher-order 
symmetry energy coefficients $C_4$, $C_6$, etc. which we do not deal here. 
The expressions for $C_2$ or the higher-order symmetry energy coefficients 
so obtained are exact within the precincts of the premises we have chosen. 
 Taking second and third density derivatives of $C_2(\rho )$ in (\ref{c2r1}),
expressions for $K_{sym}^0$ and $Q_{sym}^0$ at saturation can then be obtained, which
with the help of Eq. (\ref{c2r2}) read
\bea
\label{aksym1}
K_{sym}^0&=&-3\tilde\alpha[3C_2^0-L_0]+E_F^0 \Big [ (3\tilde\alpha -4) \nonumber \\ &&
+\left(\frac{2}{3}\frac{m}{m_0^*}+k_-\rho_0\right)(5-3\tilde\alpha ) \Big ]
\eea
and 
\bea
\label{aqsym1}
Q_{sym}^0&=&15 \tilde\alpha [3C_2^0-L_0]+K_{sym}^0(3\tilde\alpha -1) \nonumber \\ &&
+E_F^0(2-3\tilde\alpha ). 
\eea
While exploring the standard Skyrme EDFs, 
we found that exactly this correlated structure 
of $K_{sym}^0$ or $Q_{sym}^0$ with
$[3C_2^0-L_0]$ as in Eqs. (\ref{aksym1}) and (\ref{aqsym1}) is obtained. 
\begin{figure}[t]{}
\includegraphics[height=3.5in,width=3.2in,angle=-90]{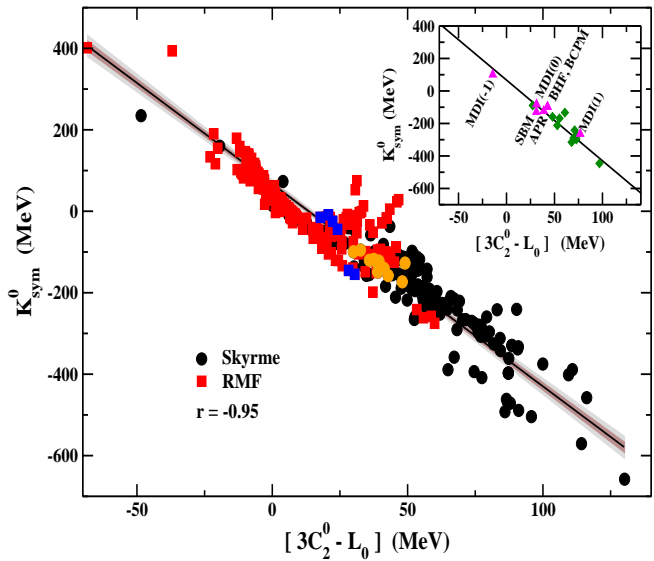}
\caption{\label{fig1}
(Color online)  The correlation between $K_{sym}^0$ and $[3C_2^0-L_0] $
as obtained from five hundred EDFs \cite{Dutra12, Dutra14}. The black circles 
correspond to
the Skyrme-inspired EDFs, the red squares refer to those obtained
from RMF models. The models consistent with all the constraints 
demanded by Dutra et al. are highlighted by orange circles for Skyrme EDFs 
\cite{Dutra12} and blue squares for RMF EDFs \cite{Dutra14}.
The inner (outer) colored regions around the best-fit straight
line  through these points depict the loci of 95$\% $ 
confidence (prediction) bands of the regression analysis. The
inset shows the correlation line obtained from the Skyrme-RMF models,
the magenta triangles are the results obtained from EDFs with
realistic interactions, MDI(0), MDI(1), MDI(-1) \cite{Chen09}, APR 
\cite{Akmal98}, BHF \cite{Taranto13}, BCPM \cite{Baldo13} and SBM 
\cite{Agrawal17}, respectively. The green diamonds represent results 
from a few Gogny interactions \cite{Sellahewa14}.} 
\end{figure}

Eq. (\ref{aksym1}) throws a  hint that there is a strong likelihood
that $K_{sym}^0$ calculated with different EDFs may be linearly correlated to
$[3C_2^0-L_0]$ corresponding to the EDFs. This is vindicated from
the correlated structure of $K_{sym}^0$ with $[3C_2^0-L_0]$
as displayed in Fig.\ref{fig1} for five hundred energy density
functionals \cite{Dutra12, Dutra14} that have been in use to explain
nuclear properties.  The results as presented in Fig. \ref{fig1}
span both the Skyrme-inspired nonrelativistic (black circles) EDFs which
tend to have negative values for $K_{sym}^0$ and also the relativistic
mean-field EDFs (red squares) that tend to have larger, sometimes positive
values for $K_{sym}^0$.  We highlight those Skyrme (orange circles) and 
RMF (blue squares) models chosen by Dutra et. al. \cite{Dutra12, 
Dutra14} which were found to satisfy specific constraints on nuclear matter 
and neutron star properties. The linear correlation as observed seems to
be nearly universal and intrinsic to an EDF consistent with nuclear
properties. The correlation coefficient ($r$) is seen to be $r=-0.95$. The
near-universality in the correlation is brought into sharper focus in
the inset in the figure where results corresponding to EDFs obtained from 
several realistic interactions (magenta triangles) and a few Gogny interactions 
(green diamonds) are displayed and are seen to lie nearly on the 
correlation line. The linear regression analysis yields
\bea
\label{corr1}
K_{sym}^0=d_1[3C_2^0-L_0] +d_2 ,
\eea
with $d_1 =-4.97 \pm 0.07$ and $d_2 =66.80 \pm 2.14$ MeV.
This is a robust correlation among the symmetry energy elements. Incidentally, 
from the density-dependent M3Y (DDM3Y) interaction,  
a similar kind of relation between these symmetry elements can be 
observed \cite{Dong12}. 
The correlation between the $K_{sym}^0$ and $L_0$ values from different effective 
forces and realistic interactions has also been considered in previous literature 
\cite{Yoshida06,Danielewicz09,Chen09,Vidana09,Ducoin11,Dong12,Santos14}. 
The results have shown relatively varying degrees of correlation. In our case, 
the correlation between $K_{sym}^0$ and $L_0$ from 
all the five hundred EDFs of Fig.~\ref{fig1} is not found to be as strong
as the correlated structure of $K_{sym}^0$ with $[3C_2^0-L_0]$, the 
correlation coefficient between $K_{sym}^0$ and $L_0$ is seen to be $r=0.87$.
The correlation between $Q_{sym}^0$ with $(3C_2^0-L_0)$ is
incidentally not as good as for $K_{sym}^0$, the correlation 
coefficient is 0.66. This is possibly because of propagation 
of errors from $K_{sym}^0$. 

From accumulated experimental data over several decades and their 
theoretical analyses, there seems to be a broad consensus about the values of 
some of the nuclear constants. The saturation density $\rho_0$ of SNM, its 
energy per nucleon $e_0$ and its incompressibility coefficient $K_0$ are 
taken as a subset of the constants characterizing symmetric nuclear matter. 
The nucleon effective mass $m_0^*$ for SNM at $\rho_0$ is also taken as an 
input datum though its value is not as certain as $e_0$ or $\rho_0$. Two 
more nuclear constants related to asymmetric nuclear matter (ANM) are 
further considered. They 
are the nuclear symmetry energy coefficients $C_2(\rho)$ at $\rho_0$ and 
at a somewhat lesser density $\rho_1$ (= 0.1 fm$^{-3}$), ``the crossing density''. 
There is less room 
for uncertainty in the symmetry energy coefficient $C_2^0$ which has been 
determined from exploration of nuclear masses \cite{Moller12, Jiang12}. 
With the realization that the nuclear observables related to average 
properties of nuclei constrain the nuclear EDFs better at around the average 
density of terrestrial atomic nuclei \cite{Khan12}, the so-called 
``crossing density'' \cite{Brown13} assumes a special significance. The 
symmetry energy $C_2^1$ ($= C_2(\rho_1)$) at that density, in Skyrme EDFs is 
seen to be strongly correlated to the Giant Dipole Resonance (GDR) in 
spherical nuclei and is now fairly well constrained \cite{Trippa08}. From 
the apparently universal, EDF-independent correlation between the isovector 
observables, the isovector elements $L_0$, $K_{sym}^0$, etc. can now be 
threaded to the above-mentioned nuclear constants as we show below.

With $m_0^*$ as input, $k_+$ is known. From given values of $e_0,\ \rho_0$ 
and $K_0$  for SNM, $\tilde\alpha$ can be  calculated as \cite{De15}, 
\bea
\label{alpha}
\tilde\alpha =\frac{\frac{K_0}{9}+\frac{E_F^0}{3}(\frac{12}{5}-2\frac{m}{m_0^*})}
{\frac{E_F^0}{5}(3-2\frac{m}{m_0^*})-e_0}.
\eea
The symmetry energy $C_2(\rho_1 )$ can be expressed as 
\bea
\label{c2r3}
C_2(\rho_1 )&=& C_2^0-L_0\epsilon+\frac{1}{2}K_{sym}^0 \epsilon^2\nonumber\\
&&-\frac{1}{6}Q_{sym}^0 \epsilon^3+\cdots,
\eea
where $\epsilon = \frac{(\rho_0 -\rho_1 )}{3\rho_0}$. 
From Eqs.~(\ref{aqsym1}), (\ref{corr1}) and (\ref{c2r3}),  
ignoring terms beyond
$\epsilon^3$,  which are negligible,
$L_0$, $K_{sym}^0$ and $Q_{sym}^0$
are calculated with known values of $C_2^0$ and $C_2^1$. 
The constant $k_-$ then follows from Eq. (\ref{aksym1}).
 From Eq. (\ref{mstar6}), the nucleon effective 
mass splitting at saturation density to leading order in $\delta$ is given 
as 
\bea
\label{mstar8}
\Delta m^*_0=\left(\frac{m_n^*-m_p^*}{m}\right)_{\rho_0}
\simeq -k_-\rho_0\left(\frac{m_0^*}{m}\right)^2\delta,
\eea
where  the approximation $ (m_n^*\cdot m_p^*)\simeq (m_0^*)^2$ is made.

Comparing Eqs. (\ref{aksym1}) and (\ref{corr1}) one would expect $|d_1|$ to be 
close to $3\tilde{\alpha}$. With the input values of the isoscalar 
nuclear constants $e_0$, $\rho_0$ and $K_0$, $3\tilde{\alpha}$ is 
seen to be 3.54 as opposed to $\sim 5$ for $|d_1|$. The reason for this 
change seems to be two-fold, (a) all 500 EDFs employed in Fig. \ref{fig1} 
have different values for $\tilde{\alpha}$, and (b) the RMF models are 
also included in the fit which have no explicit counterpart of $\tilde{\alpha}$. 

In summary, the values of $L_0$, $K_{sym}^0$, $Q_{sym}^0$ and $\Delta m_0^*$ can
be calculated in terms of empirically known nuclear constants namely,
$\rho_0$, $e_0$, $K_0$, $C_2^0$, $C_2^1$ and $\frac{m_0^*}{m}$
using Eqs.~(\ref{aqsym1})--(\ref{mstar8}).
From the diverse theoretical endeavours like the liquid drop type models 
\cite{Moller12, Myers98,Myers96}, the microscopic ab-initio or variational 
calculations \cite{Akmal97, Baldo13}, or different Skyrme or RMF models -- all initiated 
to explain varied experimental data, a representative set 
of the input nuclear constants for SNM is chosen with $\rho_0=0.155 
\pm 0.008$ fm$^{-3}$ and $e_0=-16.0\pm0.2$ MeV. From microscopic analysis 
of isoscalar giant monopole resonances (ISGMR), the value of $K_0$ 
is constrained as $230\pm40$ MeV \cite{Khan12}. Analyzing the compact 
correlation between the 'experimental' double-differences of symmetry 
energies of finite nuclei and their mass number, Jiang et. al. \cite{
Jiang12} find $C_2^0=32.1\pm0.3$ MeV. We include this value in our 
set of nuclear constants. For $C_2^1$, the value $C_2^1=24.1\pm0.8$ MeV
as quoted from microscopic analysis of GDR in $^{208}$Pb \cite{
Trippa08} is taken. There is an overall consistency of this $C_2^1$ value with those 
from the best-fit Skyrme EDFs \cite{Brown13} and with that given in 
\cite{Dong12}.
For the nucleon effective mass, a value of $\frac{m_0^*}{m} =0.70 \pm 0.05
$ is taken, this is consistent with the empirical values obtained from
many analyses \cite{Jaminon89,Li15b}.

The values of the symmetry energy elements calculated from Eqs.~(\ref{aqsym1})--(\ref{mstar8}) 
using the values of input nuclear constants as mentioned
come out to be $L_0=60.3\pm14.5$ MeV, $K_{sym}^0=-111.8\pm71.3$
MeV, $Q_{sym}^0 =296.8\pm73.6$ MeV and $\Delta m^*_0=(0.17\pm0.24)
\delta $. The value of $L_0$ is remarkably close to its global average
58.9$\pm $ 16 MeV \cite{Li13}, obtained from analyses of terrestrial
experiments and astrophysical observations. The value of $L$ at $\rho_1$ 
is calculated to be $49.3\pm4.2$ MeV. From dipole polarizability in 
$^{208}$Pb an empirical value of $L=47.3\pm7.8$ MeV was obtained 
at $\rho\simeq0.11$ fm$^{-3}$ \cite{Zhang14a}.
To our knowledge, there is no
experimental value for $K_{sym}^0$ or $Q_{sym}^0$ to compare. However, the
symmetry incompressibility $K_\delta$ defined at the saturation density of
nuclear matter at asymmetry $\delta $ ($K_\delta =K_{sym}^0 -6L_0 -\frac
{Q_0 L_0}{K_0}$, where $Q_0=27\rho_0^3\left(\frac{\partial^3e}{\partial
\rho^3}\right)_{\rho_0}$) has been extracted from breathing mode energies
of Sn-isotopes \cite{Li07}. Corrected for the nuclear surface term,
$K_\delta$ is quoted to be $\simeq -350$ MeV \cite{Pearson10}. This is in
close agreement with our calculated value $K_\delta =-378.6\pm17.0$ MeV; $Q_0$ has
been calculated from Eq.~(\ref{ener4}) to be $-364.7\pm27.7$ MeV corresponding
to $\delta =0 $ \cite{De15} with the input nuclear constants mentioned.
Experimental search for $\Delta m^*_0$, till now is scanty and not very
certain. From analysis of nucleon-nucleus scattering data within an
isospin dependent optical model \cite{Li15b} it is reported to be (0.41
$\pm $0.15)$\delta $. From exploration of isoscalar giant quadrupole
resonance and dipole polarizability \cite{Zhang16} in $^{208}$Pb it goes
down to (0.27$\pm $0.15)$\delta $. Our calculated value for $\Delta
m^*_0$  is on the smaller side. In a recent dynamical BUU calculation 
\cite{Kong17} with improved isospin and momentum dependent interaction 
where the isovector giant dipole resonance properties of $^{208}$Pb were 
used to constrain the nuclear symmetry energy slope parameter and the isospin 
splitting of the nucleon effective mass, $\delta m_0^*$ comes close to our 
value. As a test of the viability of the methodology 
we have used in this communication, we also calculated $L_0$, $K_{sym}^0$ 
and $Q_{sym}^0$ for the 16 Skyrme models selected through several 
tests by Dutra et. al. \cite{Dutra12} using their model values of $\rho_0$, 
$e_0$, $K_0$, $C_2^0$, $C_2^1$ and $\frac{m_0^*}{m}$ as inputs. 
The so-obtained values of $L_0$, $K_{sym}^0$ and $Q_{sym}^0$ come out 
to be pretty close to their model values given in Ref. \cite{Dutra12} 
with an average root mean square deviation of 1.3 MeV in $L_0$, 21.4 MeV in 
$K_{sym}^0$ and 32.5 MeV in $Q_{sym}^0$, showing the reasonableness of 
our adopted method.

\begin{figure}[t]{}
\includegraphics[height=4.0in,width=3.5in]{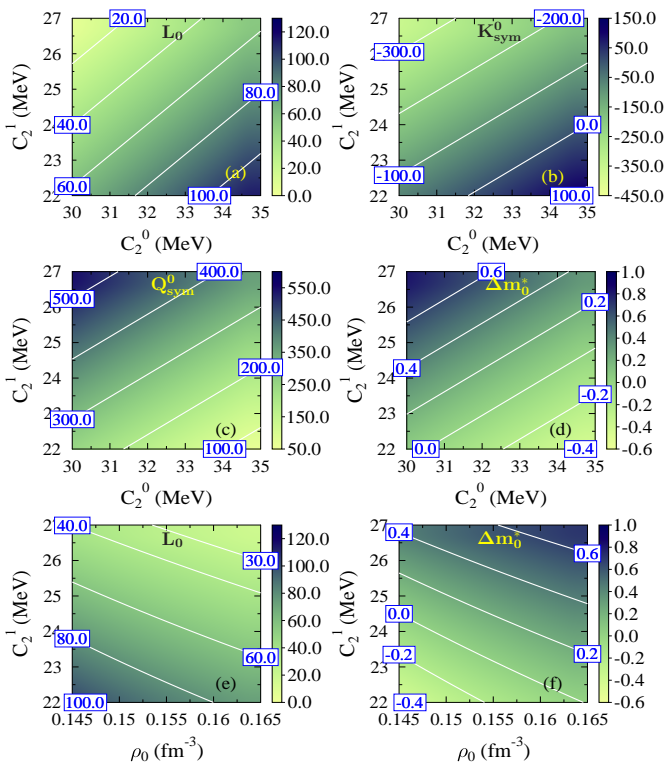}
\caption{\label{fig2}
(Color online)  Contours of constant $L_0$, $K_{sym}^0$, $Q_{sym}^0$
and $\Delta m_0^*$ in color shades (as indicated on the right side
of each panel) as functions of the input nuclear constants $C_2^0$,
$C_2^1$ and $\rho_0$ depicting the interdependence between various 
symmetry energy elements. The values of $L_0$, $K_{sym}^0$ and 
$Q_{sym}^0$ are in units of MeV and those for $\Delta m_0^*$ are 
in units of the free nucleon mass. For details, see text.} 
\end{figure}

The set of nuclear constants we have chosen is a conservative set; 
depending on possible new experimental inputs, their values may however change 
somewhat which would affect the calculated values of the density derivatives 
of the symmetry energy coefficients. The evaluated isovector elements
are seen to be quite sensitive to the input quantities $C_2^0$,
$C_2^1$ and $\rho_0$. There is still some variance in the choice of these 
input nuclear constants \cite{Zhang13,Gang08,Liu10} besides the ones we have chosen. The
aforesaid sensitivity can be gauged from the
displayed six panels in Fig. \ref{fig2}. In the upper four panels
(a)-(d), the contours of constant $L_0$, $K_{sym}^0$, $Q_{sym}^0$
and $\Delta m_0^*$ are shown in the $C_2^0-C_2^1$ plane
in color shades, the white lines within the panels are the loci of
constant isovector elements as marked when all other input elements are
left unchanged.  With increase in   $C_2^1$, $L_0$ and $K_{sym}^0$
are seen to decrease whereas $Q_{sym}^0$ and $\Delta m_0^*$ are found to
increase. The opposite is observed for an increase in $C_2^0$. This
points out the interdependence between different symmetry
energy elements. The change in $\rho_0$ has also a sizeable effect on the
isovector elements.  All other inputs remaining intact, an increase in
$\rho_0$ decreases $L_0$ and $K_{sym}^0$ and increases $Q_{sym}^0$ and
$\Delta m_0^*$.  We show only glimpses of these changes in panels (e)
and (f), where contours of constant $L_0$ and $\Delta m_0^*$ are drawn
in the $\rho_0-C_2^1$ plane. The isovector elements as studied
here are seen to be nearly insensitive to changes in $e_0$ and $m_0^*$
(not shown here). Similarly, $K_0$ has little effect on these isovector
elements except on $\Delta m_0^*$.  An increase of $K_0$ by, e.g., $\sim $
30 MeV is seen to push $\Delta m_0^*$ drastically in the negative domain.
Uncertainties in the input nuclear constants bear signature on
the uncertainties in the calculated isovector elements.

To sum up, without  reference to any specific nuclear interaction,
with only a few reasonable approximations, 
analytic expressions for the density derivatives
of the symmetry energy coefficient $C_2(\rho )$ at the saturation density in
terms of  empirical nuclear constants are found out. 
The symmetry observables are seen to be sensitive to the values of
the input nuclear constants, particularly to $C_2^0$, $C_2^1$ and
$\rho_0$; precise values of these constants are thus required to
narrow down the uncertainties in the density dependence of the symmetry
energy. In doing the calculations, a correlated structure connecting 
the different symmetry energy elements emerged. The
consonance of these structures with those inherent in the
plethora of EDFs of different genre used in nuclear microscopy is
very revealing. This indicates a universality in the correlated
structure in the symmetry energy coefficients and helps in a better  
realization of the information content of the isovector observables.

\acknowledgments
J.N.D. is thankful to the Department of Science and Technology, Government of India 
for support with the Grant EMR/2016/001512.
M.C. and X.V. acknowledge support from Grant FIS2014-54672-P from MINECO and FEDER, 
Grant 2014SGR-401 from Generalitat de Catalunya, and the project MDM-2014-0369 of ICCUB 
(Unidad de Excelencia Mar\'{\i}a de Maeztu) from MINECO.


\end{document}